\documentclass[a4]{epl2}
\usepackage{graphicx}
\usepackage{epsfig}
\usepackage{color}
\usepackage{graphicx}
\usepackage{epstopdf}
\usepackage{amsmath}
\usepackage{amsfonts}
\usepackage{bm}
\usepackage{float}

\usepackage{fancyhdr}
\title{\textcolor[rgb]{0.00,0.00,1.00}{Quantum mechanics with quaternionic mass}}

\author{A. I. Arbab\inst{}\footnote{arbab.ibrahim@gmail.com}}
\date{today}
\institute{
Department of Physics, College of Science, Qassim University, P.O. Box 6644,  Buraidah 51452,  Saudi Arabia}


\abstract{ Quantum mechanics with quaternionic mass is considered. The momentum eigen-value equation with quaternionic mass yields the Klein-Gordon equation with a mass consisting of longitudinal and traverse masses. The scalar field total mass is found to be a sum of these  masses. This field appears to  be connected with two subfields conserving linear momentum.  It is found that a particle with real mass satisfies the quantum Telegraph equation, whereas that one with quaternionic mass satisfies the Klein-Gordon equation.  A quantum force acting on the particle is found to be proportional to its velocity. When the particle field is coupled to an electromagnetic field, an additional term in the particle's energy appears reflecting the interaction of the particle's angular momentum with the magnetic field.}

\begin{document}
\date{today}
\maketitle
\baselineskip=18pt

\section{Introduction}

 Maxwell, Planck, de Broglie, and Einstein have laid down the foundational principles  of the elementary particles and their fields \textcolor[rgb]{0.00,0.00,1.00}{\cite{bjorken}}.   In Maxwell's theory, the electric and magnetic fields are treated as one entity if the theory is invariant under the duality transformation. Under such circumstances,  Maxwell equations are reduced to two equations.  Einstein states that energy and mass are equivalent so that each one can be transformed into another.   Lorentz had found that space and time are one entity. They can be transformed into each other by employing Lorentz transformations. Since matter is inscribed into mass and energy into a wave, then matter and wave are equivalent too. This is normally coined in the wave-particle duality. In quaternionic jargon, the quantum equation reflecting the wave-particle duality  is the momentum eigen-value equation. This yields a new quantum wave equation exhibiting this duality aspect \textcolor[rgb]{0.00,0.00,1.00}{\cite{epl,qqm}}.

The mathematical formalism in which the above principles are realized is via equations of motion. At their highest level, these are the Maxwell, Klein-Gordon, and Dirac equations. Maxwell equations describe pure wave phenomena, and Klein-Gordon and Dirac describe the particle phenomena.  However an equation that describes the evolutionary state is found in  \textcolor[rgb]{0.00,0.00,1.00}{\cite{epl,qqm}}. The two forms of existence are described by fields. Any change in this field will be exhibited as a wave. Besides the dynamical evolution of a quantum object, an aggregation of particles in an ensemble is found to be governed by specific statical laws. Further, it is found that the dynamics and  statistics are connected. Fermions that are governed by the Dirac equation follow different statistics than those of bosons which are described by the Klein-Gordon equation. Besides, it is  also found that  Maxwell equations govern another kind of bosons with different statistics. In effect, all particles find their way to be described by appropriate dynamical as well as statistical equations.

 Objects like molecules (atoms) follow other statistics known as the Maxwell-Boltzmann statistics. Therefore, the physical properties of microparticles are intimately connected.  It is well known that all objects (particles)  have inertia expressing the extent to which they resist a change of state (motion to rest or rest to motion). Inertia can be modeled as some kind of friction. In quantum mechanics, a quantum particle is modeled by a wave packet in such a way that its group velocity is equal to the particle velocity. However, the phase velocity of a single wave is different from the particle and group velocities.  In our anticipated description, we think of a particle and its wave as two manifestations of one object. When the particle behaves as a wave its momentum is equal to that when it behaves like a normal particle. The wave has a vector mass whereas the particle has a scalar mass. The two entities are connected, however.

Space can not be compatible with the notion of inertia, however. A fluid (ether) filling the whole space was promoted by physicists but later rejected by the famous Michelson–Morley experiment. The experiment did not rule out the presence of the fluid but rather excluded its existence to carry the electromagnetic wave. The electromagnetic wave was thought of as a mechanical wave that requires a priori a medium to propagate in. The modern understating of the space, today, allows some kind of minimal field to be pervading the whole space. This expresses a background due to quantum fluctuations in the field. It is called the \emph{vacuum} that today bears a physical existence.

Mass is a property that distinguishes an object.  The wave inertia however is not characterized by its mass. Therefore, the notion of mass has to be extended to incorporate particles as well as waves. There exists now a need to distinguish different kinds of mass that could be due to the field of force under which the object interacts. With the same token, an object can have two kinds of  charge too playing analogous roles. One mass (charge) represents the matter nature while  the other one characterizes the wave nature. A mass besides being the source of matter-wave could also act as a source for the electromagnetic field.

 We have found that  Quaternions are the appropriate mathematical tools that can deal with all of the above identification of the object. This is owing to the spectacular algebra of Quaternions. Treating the particle mass as  quaternion yields very interesting results. Since a quaternion consists of scalar and vector components, a quaternionic mass embodies the particle (scalar) and wave (vector) aspects. In the de Broglie theory,  a photon wave, with zero mass, is used to draw a parallel with that one having a mass.

 In the present formalism, we however employ a different prescription.   We would like to explore these possibilities  by applying the quaternionic mass in Dirac formalism to manifest this. This recipe turned out to be equivalent to introducing an interaction  of the quantum particle with an external field. Previously,  the  normal mechanism to do so is via the minimal coupling method.

 This paper is scheduled as follows: we introduce in Section 2 quaternionic quantum mechanics with real and quaternionic mass. In this section, we derived the energy conservation (continuity) equation of the quantum system. We then showed that the duality aspect is guaranteed by deriving a particle-wave momentum equation connecting the wave momentum to that of the particle. This relationship reveals that a quaternionic mass representation of a quantum particle realizes the duality principle due to de Broglie. While the ordinary quaternionic momentum eigen-value equation with real mass yields a Telegraph equation, that one with quaternionic mass yields Klein-Gordon equation for a field whose mass consists of two masses. In Section 3 we study the interaction of a quantum particle with an electromagnetic field. An interaction term showing the coupling of a particle's angular momentum with the magnetic field appears in the particle's Hamiltonian. A Lorentz-like force deemed to govern the quantum particle motion is presented in Section 4. We end our paper by concluding remarks in Section 5.

\section{Quantum mechanics with quaternionic mass}

The Dirac (momentum eigen-value) equation with real mass, $m$, was studied before yielding \textcolor[rgb]{0.00,0.00,1.00}{\cite{epl,qqm}}
\begin{equation}
\frac{1}{c^2}\,\frac{\partial\psi_0}{\partial t}+\vec{\nabla}\cdot\vec{\psi}+\frac{m}{\hbar}\,\psi_0=0\,,
\end{equation}
\begin{equation}
  \frac{\partial\vec{\psi}}{\partial t}+\vec{\nabla}\psi_0+\frac{mc^2}{\hbar}\,\vec{\psi}=0\,\,,
\end{equation}
\begin{equation}
\vec{\nabla}\times\vec{\psi}=0\,.
\end{equation}
Equations (1) - (3) yield the wave equations
\begin{equation}
\frac{1}{c^2}\frac{\partial^2\psi_0}{\partial t^2}-\nabla^2\psi_0+\frac{2m}{\hbar}\frac{\partial\psi_0}{\partial t}+\left(\frac{mc}{\hbar}\right)^2\psi_0=0\,,\qquad \frac{1}{c^2}\frac{\partial^2\vec{\psi}}{\partial t^2}-\nabla^2\vec{\psi}+\frac{2m}{\hbar}\frac{\partial\vec{\psi}}{\partial t}+\left(\frac{mc}{\hbar}\right)^2\vec{\psi}=0\,.
\end{equation}
These are the undistorted Telegraph equations that are known to describe the propagation of the electric current and voltage in transmission lines. Interestingly, the inertial (quantum) and electromagnetic waves of the electron are coherent. The solution of the above quantum Telegraph equation is  a wavepacket consisting of waves traveling to the left and right with the speed of light. This is an advantage of the quantum Telegraph equation over the Schrodinger equation where the solution of the latter  is a plane wave. For the Schrodinger equation to describe the motion of a particle  two plane waves have to be added to obtain a wavepacket solution.

We would like here to entertain the possibility of studying the momentum eigen-value equation with  quaternionic mass. To this aim let us write
the quaternionic momentum eigen-value generalizing the ordinary one. It is expressed in the form \textcolor[rgb]{0.00,0.00,1.00}{\cite{epl,qqm}}
\begin{equation}
\tilde{P}\,\tilde{\Psi}^*=c\tilde{M}\,\tilde{\Psi}^*\,,\qquad \tilde{P}=\left(\frac{i}{c}\,\hat{E}\,,\hat{\vec{p}}\,\right)\,, \qquad \tilde{\Psi}=\left(\frac{i}{c}\,\psi_0\,,\vec{\psi}\,\right)\,,\qquad  \tilde{M}=\left(im_\ell\,,\vec{m}_t\,\right)\,
\end{equation}
 Here, $\hat{E}$ and $\hat{\vec{p}}$ are the energy and momentum operators. The two masses $m_\ell$ and $\vec{m}_t$  are the longitudinal and transverse masses describing the wave-particle duality. Here $\tilde{\Psi}$ describes the state of the quantum particle consisting of a vector and a scalar part that is analogous to the scalar and vector potentials describing the electromagnetic field. The quaternion mass, $\tilde{M}$, can be thought of as a mass field. The particle interacts with the outside world by coupling this mass field with the other fields in which the particle exists. The physical meaning of the masses, $\vec{m}_t$ and $m_\ell$  will be given later on.

 Applying the quaternionic product rule, and equating the real and imaginary parts in the two sides of the resulting equations, one finds  \textcolor[rgb]{0.00,0.00,1.00}{\cite{hisham}}
\begin{equation}
\frac{1}{c^2}\,\frac{\partial\psi_0}{\partial t}+\vec{\nabla}\cdot\vec{\psi}=0\,,
\end{equation}
\begin{equation}
  \frac{\partial\vec{\psi}}{\partial t}+\vec{\nabla}\psi_0+\frac{c^2}{\hbar}\,\vec{m}_t\times\vec{\psi}=0\,\,,
\end{equation}
\begin{equation}
\vec{\nabla}\times\vec{\psi}+\hbar^{-1}(m_\ell c\,\vec{\psi}-\, \vec{m}_t\psi_0)=0\,,
\end{equation}
and
\begin{equation}
m_\ell\psi_0-c\,\vec{m}_t\cdot\vec{\psi}=0\,.
\end{equation}
Now equations (7) and (9) yield
\begin{equation}
\frac{\partial \rho}{\partial t}+\vec{\nabla}\cdot\vec{J}=0\,,\qquad \vec{J}=c\,\vec{m}\psi_0\,,\qquad \rho=m_\ell \psi_0\,.
\end{equation}
Thus if $\vec{J}=\rho\,\vec{v}$, then
\begin{equation}
m_\ell\vec{v}=c\,\vec{m}_t\,.
\end{equation}
Equation (11) can be seen as momentum conservation of the two masses representing the quantum particle. Since in de Broglie's theory a particle has a wave nature, the term $c\,\vec{m}_t$ in eq.(11) could account for  the momentum of the wave associated with the particle. If we say that the matter-wave momentum is $\vec{p}_W=c\vec{m}_t$, then using Einstein's mass-energy relationship, $E=mc^2$, one finds the value of the matter-wave momentum to be $p_W=E/c$\,.  This agrees with the standard representation of the momentum of light (photon). The vector mass, $\vec{m}_t$,  points along the direction in which matter-wave travels. However, in the standard quantum mechanics, a particle is represented by a wavepacket that travels at a velocity (group velocity) equals to the particle's velocity.

Recall that when a charged particle interacts with an electromagnetic field, the field imparts  momentum $\vec{p}_A=q\vec{A}$ such that the total momentum the charge receives is $\vec{p}\,'=\vec{p}+q\vec{A}$. This allows us to deduce that the  momentum provided to the particle is $\vec{p}_f=c\,\vec{m}_t$. When a charged particle interacts with an electromagnetic field then, $q\vec{A}=c\,\vec{m}_t$. Hence, the transverse (vector) mass describes the momentum that a particle receives while moving. If the particle moves freely, $\vec{m}_t=0$. This urges us to define $m_\ell$ as the mass of the free particle. It is thus very intriguing to encapsulate the two masses in a single representation, the quaternionic mass.

Now eqs.(6) and (7) yield
\begin{equation}
\frac{\partial u}{\partial t}+\vec{\nabla}\cdot\vec{S}=0\,,\qquad\qquad \vec{S}=\psi_0\vec{\psi}\,,\qquad\qquad u=\frac{\psi^2}{2}+\frac{\psi_0^2}{2c^2}\,,
\end{equation}
which represents the energy conservation equation of the system. Here $u$ represents the energy density of the particle field and $\vec{S}$ is its energy flux. The quantum particle/de Broglie wave travels/propagates in space is like the fluid flow. It was Maxwell who thought of electromagnetic field (wave) propagation as fluid flow. It is interesting to observe that the introduction of the quaternionic mass removes the dissipation in the system that is associated with eqs.(1) - (3).

Multiply eq.(7) by $\psi_0$ and employ eqs.(6) and (8) to get the momentum (force) equation
\begin{equation}
\frac{1}{c^2}\,\frac{\partial\vec{S}}{\partial t}+(\vec{\psi}\vec{\nabla}\cdot\vec{\psi}+\vec{\psi}\cdot\vec{\nabla}\vec{\psi})+\vec{\nabla}\left(\frac{\psi_0^2}{2c^2}-\frac{\psi^2}{2}\right)=0\,,
\end{equation}
In components form,  eq.(13) can be expressed as
\begin{equation}
\frac{1}{c^2}\,\frac{\partial S_i}{\partial t}+\partial_j (\psi_i\psi_j)+\partial_i\left(\frac{\psi_0^2}{2c^2}-\frac{\psi^2}{2}\right)=0\,,
\end{equation}
which can be written as
\begin{equation}
\frac{1}{c^2}\,\frac{\partial S_i}{\partial t}+\partial_j (\psi_i\psi_j+\delta_{ij}(\frac{\psi_0^2}{2c^2}-\frac{\psi^2}{2}))=0\,,\qquad-\frac{\partial g_i}{\partial t}+\partial_j \sigma_{ij}=0\,,
\end{equation}
where
\begin{equation}
\sigma_{ij}=-\psi_i\psi_j-\delta_{ij}\left(\frac{\psi_0^2}{2c^2}-\frac{\psi^2}{2}\right)\,,
\end{equation}
is a symmetric stress-tensor of the matter field. Interestingly, eq.(15) is the momentum conservation equation that is analogous to the electromagnetic, where the momentum of the matter wave is defined by $\vec{g}=\frac{\vec{S}}{c^2}$. Thus,  eq.(15) describes the force density that the matter wave impinges. It is similar to Euler's equation of fluid motion. This indicates that the motion of a quantum particle is analogous to fluid motion.

Eqs.(6) - (9) can be shown to represent a wave equation (matter wave) \textcolor[rgb]{0.00,0.00,1.00}{\cite{hisham}}
\begin{equation}
\frac{1}{c^2}\,\frac{\partial^2\psi_0}{\partial t^2}-\nabla^2\psi_0+\left(\frac{M c}{\hbar}\right)^2\psi_0=0,\qquad \frac{1}{c^2}\,\frac{\partial^2\vec{\psi}}{\partial t^2}-\nabla^2\vec{\psi}+\left(\frac{M c}{\hbar}\right)^2\vec{\psi}=0\,,
\end{equation}
where
\begin{equation}
M^2=m_t^2- m_\ell^2\,.
\end{equation}
Equation (17) is but the Klein-Gordon equation of a scalar field whose effective mass is $M$ which I prefer to call  the invariant mass. It is thus interesting that the momentum eigen-value equation with a quaternion mass gives the Klein-Gordon equation. It seems that the Klein-Gordon field is a superposition of two kinds of fields. A tachyonic Klein-Gordon equation results when $m_t<|m_\ell|$. As evident from eq.(18), if the scalar field has the same mass components, it then behaves like a massless field (photon). It is quite remarkable to see that a particle described by a real mass satisfies the quantum Telegraph equation while that one described by a quaternionic mass yields the Klein-Gordon equation. Note that the Dirac equation can be obtained from the quantum Telegraph equation by taking the mass to be imaginary \textcolor[rgb]{0.00,0.00,1.00}{\cite{epl}}.

The quaternionic momentum eigen-value equation associated with real mass was found to lead to the Telegraph equation (or a dissipative Klein-Gordon equation) \textcolor[rgb]{0.00,0.00,1.00}{\cite{qte}}. The dissipative term is removed when quaternionic mass is proposed instead.

 Applying eq.(11) in eq.(18) yields
\begin{equation}
m_\ell=\frac{M}{\sqrt{1-\frac{v^2}{c^2}}}\,,\qquad\qquad \vec{m}_t=\left(\frac{\vec{v}}{c}\,\right)m_\ell\,.
\end{equation}
The longitudinal mass $m_\ell$ is but the relativistic mass and $M$ is the field rest-mass. Notice that in general the value of the transverse mass is smaller than the longitudinal one. We could now say that the transverse mass is the \emph{active mass} whereas the longitudinal mass is a \emph{passive mass} of the particle. When the particle is not interacting its transverse vanishes and its longitudinal mass remains. This new concept of masses should find application in scattering problems of microscopic particles having particle and wave natures. In such cases, the conservation of energy and momentum should be taken in the association of a quaternionic perspective.

Interestingly,  Eq.(19) associates inertia with motion. In the ordinary Newton's laws, the inertia of a macroscopic object is connected with its acceleration. However, in quantum mechanics the force is directly probational to  velocity  \textcolor[rgb]{0.00,0.00,1.00}{\cite{q-force}}. The second equation in eq.(19) can be expressed as the Einstein's relativistic equation between velocity, momentum and energy that is valid for matter as well as waves, $\vec{v}=(c^2/E)\,\vec{p}$, upon expressing it as  $\vec{v}=[c^2/(m_\ell c^2)]\vec{m}_tc$\,, with $E=m_\ell c^2$ and $\vec{p}=\vec{m}_tc$\,. The latter momentum can be seen as a  momentum of the wave-packet mimicking the particle. Hence, this analogy provides a faithful realization of our theory. Thus, for particle interactions the momentum and energy conservations are conservations of the vector mass and relativistic masses, $\vec{m}_t$ and $m_\ell$\,, respectively. The relation in eq.(19) was already derived in the context of a different eigen-value equation of a quantum particle \textcolor[rgb]{0.00,0.00,1.00}{\cite{arbab et al}}.

Defining $\rho=\psi^2_0/c^2$ and $\vec{J}=\vec{\psi}$, equations (6) and (7) can be seen as the continuity and momentum conservation equations governing matter flow. It is valid for mass as well as charge flow. It therefore fully  characterizes the moving quantum particle.

\section{Interacting quantum particle with an electromagnetic field}

The interaction of a charged ($q$) quantum particle with an electromagnetic field can be found by using the minimal coupling recipe, where the particle momentum and energy are replaced, respectively,  by, $\vec{p}\rightarrow \vec{p}-q\vec{A} $ and $E\rightarrow E-q\varphi$  \textcolor[rgb]{0.00,0.00,1.00}{\cite{bjorken}}. This is found to alter eqs.(1) - (3) to read \textcolor[rgb]{0.00,0.00,1.00}{\cite{matter}}
\begin{equation}
\frac{1}{c^2}\,\frac{\partial\psi_0}{\partial t}+\vec{\nabla}\cdot\vec{\psi}+\frac{m}{\hbar}\,\psi_0=0\,,
\end{equation}
\begin{equation}
  \frac{\partial\vec{\psi}}{\partial t}+\vec{\nabla}\psi_0+\frac{mc^2}{\hbar}\,\vec{\psi}+\frac{cq}{\hbar}\,\vec{A}\times\vec{\psi}=0\,\,,
\end{equation}
\begin{equation}
\vec{\nabla}\times\vec{\psi}+\hbar^{-1}(c^{-1}q\varphi\,\vec{\psi}-c^{-1}q\vec{A}\psi_0)=0\,,
\end{equation}
and
\begin{equation}
c^{-2}q\varphi\,\psi_0-q\vec{A}\cdot\vec{\psi}=0\,,
\end{equation}
where $\varphi$ and $\vec{A}$ are the scalar and vector potentials of the electromagnetic field.
An interesting analogy between eqs.(20) - (23) and eqs.(6) - (9) can be drawn. This is as follows: $m=0$, $c\,\vec{m}_t\rightarrow q\vec{A}$, and $m_\ell c^2\rightarrow q\,\varphi$.  Therefore, introducing a quaternionic mass in a theory is equivalent to dealing with interaction of a particle with an external field.

Equations (20) - (23) reveal that $\psi_0$ satisfies the Telegraph's wave equation
\begin{equation}
\frac{1}{c^2}\frac{\partial^2\psi_0}{\partial t^2}-\nabla^2\psi_0+\frac{2m}{\hbar}\frac{\partial\psi_0}{\partial t}+\left(\frac{m^2c^2}{\hbar^2}+\frac{q^2}{\hbar^2}(A^2-\frac{\varphi^2}{c^2})\right)\psi_0=\frac{cq}{\hbar}\vec{\psi}\cdot\vec{B}\,,
\end{equation}
and
\begin{equation}
\frac{1}{c^2}\frac{\partial^2\vec{\psi}}{\partial t^2}-\nabla^2\vec{\psi}+\frac{2m}{\hbar}\frac{\partial\vec{\psi}}{\partial t}+\left(\frac{m^2c^2}{\hbar^2}+\frac{q^2}{\hbar^2}(A^2-\frac{\varphi^2}{c^2})\right)\vec{\psi}=\frac{q}{c\hbar}(\vec{E}\times\vec{\psi}+\psi_0\vec{B})\,,
\end{equation}
with  variable mass and source terms that couple to the magnetic field. It is intriguing that the  above two equations are coupled to each other. The right-hand sides of eqs.(24) and (25) show that the electromagnetic field couple to the particle's fields (wavefunctions), $\psi_0$ and $\vec{\psi}$\,.
For homogenous wave equations, eqs.(24) and (25) imply that
\begin{equation}\tag{24a}
\vec{\psi}\cdot\vec{B}=0\,,\qquad\qquad \vec{B}=\frac{\vec{\psi}}{\psi_0}\times\vec{E}\,.
\end{equation}
These show that the electromagnetic field is perpendicular to the particle's vector wavefunction, $\vec{\psi}$\,. Equation (24a) reduces to the electrodynamics of a moving charge for, $\vec{v}\,\psi_0=c^2\vec{\psi}$\,. With the conditions in eq.(24a), eqs.(24) and (25) decouple.

It would be interesting if we connect the vector wavefunction to the angular momentum, $\vec{L}$, of the particle by
 \begin{equation}
 c\hbar\,\vec{\psi}=\vec{L}\psi_0\,,
 \end{equation}
so that the right-hand side of eq.(25) will describe a particle's orbital interaction with the magnetic field of the form
\begin{equation}
\frac{1}{c^2}\frac{\partial^2\psi_0}{\partial t^2}-\nabla^2\psi_0+\frac{2m}{\hbar}\frac{\partial\psi_0}{\partial t}+\left(\frac{m^2c^2}{\hbar^2}+\frac{q^2A^2}{\hbar^2}-\frac{q^2\varphi^2}{\hbar^2c^2}-\frac{q}{\hbar^2}\vec{L}\cdot\vec{B}\right)\psi_0=0\,.
\end{equation}
Equation (27) can be expressed in an energy equation   as
\begin{equation}
\frac{\hbar^2}{2mc^2}\frac{\partial^2\psi_0}{\partial t^2}-\frac{\hbar^2}{2m}\nabla^2\psi_0+\hbar\frac{\partial\psi_0}{\partial t}+\left(\frac{mc^2}{2}+\frac{q^2A^2}{2m}-\frac{q^2\varphi^2}{2mc^2}-\frac{q}{2m}\vec{L}\cdot\vec{B}\right)\psi_0=0\,.
\end{equation}
Under the rotation of space and time coordinates by an angle of $-\pi/2$, \emph{i.e.,}  ($t\rightarrow -it$ and $r\rightarrow -ir$) \textcolor[rgb]{0.00,0.00,1.00}{\cite{compl}}, eq.(28) becomes
\begin{equation}\tag{28a}
i\hbar\frac{\partial\psi_0}{\partial t}=\frac{\hbar^2}{2mc^2}\frac{\partial^2\psi_0}{\partial t^2}-\frac{\hbar^2}{2m}\nabla^2\psi_0-\left(\frac{mc^2}{2}+\frac{q^2A^2}{2m}-\frac{q^2\varphi^2}{2mc^2}-\frac{q}{2m}\vec{L}\cdot\vec{B}\right)\psi_0\,.
\end{equation}
The last term in the left-hand side of eq.(28a) embodies  the interaction of the particle's angular momentum with the external magnetic field. The first term in eq.(28a) will have a small contribution in the low-frequency limit. For such  a case eq.(28a) approximates to Schrodinger equation. If we chose in eq.(26) the spin angular momentum instead of the orbital angular momentum, then $c\hbar\,\vec{\psi}=2\vec{S}\psi_0$, where $2\vec{S}=\vec{\sigma}$ are Pauli's matrices, then the  energy term in eq.(28a) will describe the particle's spin angular momentum. Notice that the vector wavefunction equation, eq.(25), doesn't involve such a term.

For massive electrodynamics, one multiplies eq.(25) by $\frac{\varepsilon_0c\hbar}{q}$ and choose, $\vec{\psi}=\vec{A}$ and $\psi_0=\varphi$, so that  eq.(25) becomes
\begin{equation}
\frac{1}{c^2}\frac{\partial^2\vec{Y}}{\partial t^2}-\nabla^2\vec{Y}+\frac{2m}{\hbar}\frac{\partial\vec{Y}}{\partial t}+\left(\frac{m^2c^2}{\hbar^2}+\frac{q^2}{\hbar^2}(A^2-\frac{\varphi^2}{c^2})\right)\vec{Y}=\vec{s}\,,
\end{equation}
where
\begin{equation}
\vec{s}=\varepsilon_0(\vec{E}\times\vec{A}+\varphi\vec{B})\,,\qquad\qquad \vec{Y}=\frac{\varepsilon_0c\hbar}{q}\vec{A}\,.
\end{equation}
The quantity, $\vec{s}$, is the spin density of the electromagnetic field and the field $\vec{Y}$ is a linear momentum vector field \textcolor[rgb]{0.00,0.00,1.00}{\cite{griffth}}. This later momentum is new to electrodynamics. Equation (29) shows that the spin $\vec{s}$ is the source for the momentum vector field, $\vec{Y}$\,.  Equation (27) will yield a nonlinear Telegraph's equation
\begin{equation}
\frac{1}{c^2}\frac{\partial^2\varphi}{\partial t^2}-\nabla^2\varphi+\frac{2m}{\hbar}\frac{\partial\varphi}{\partial t}+\left(\frac{m^2c^2}{\hbar^2}+\frac{q^2A^2}{\hbar^2}-\frac{q}{\hbar^2}\vec{L}\cdot\vec{B}\right)\varphi-\frac{q^2}{\hbar^2c^2}\,\varphi^3=0\,.
\end{equation}
Equation (31) can be expressed as an energy equation as
\begin{equation}
\frac{\hbar^2}{2mc^2}\frac{\partial^2\varphi}{\partial t^2}-\frac{\hbar^2}{2m}\nabla^2\varphi+\hbar\,\frac{\partial\varphi}{\partial t}+\left(\frac{mc^2}{2}+\frac{q^2A^2}{2m}-\frac{q}{2m}\vec{L}\cdot\vec{B}\right)\varphi-\frac{q^2}{2mc^2}\,\varphi^3=0\,.
\end{equation}
Upon space and time rotation by $-\pi/2$ ($t\rightarrow -it$ and $r\rightarrow -ir$) \textcolor[rgb]{0.00,0.00,1.00}{\cite{compl}}, eq.(32) yields
\begin{equation}
i\hbar\,\frac{\partial\varphi}{\partial t}=\frac{\hbar^2}{2mc^2}\frac{\partial^2\varphi}{\partial t^2}-\frac{\hbar^2}{2m}\nabla^2\varphi-\left(\frac{mc^2}{2}+\frac{q^2A^2}{2m}-\frac{q}{2m}\vec{L}\cdot\vec{B}\right)\varphi+\frac{q^2}{2mc^2}\,\varphi^3\,.
\end{equation}
Equation (33) can be seen as a late-time nonlinear Schrodinger equation of a charged particle interacting with an electromagnetic field with $\varphi$ as a wavefunction. The fifth term on the left-hand side of eq.(33) would represent the interaction orbital angular momentum with the magnetic field, where the particle spin $\vec{S}=0$\,.

\section{Extended Lorentz force}
The Lorentz force describing the force acting of the quantum particle can be extended to read
\begin{equation}
\tilde{F}=c\tilde{M}^*\tilde{V}\tilde{\nabla}\tilde{\Psi}=\left(\frac{i}{c}\, P\,, \vec{f}\right),\qquad\qquad \tilde{V}=(ic\,,\vec{v})\,,
\end{equation}
where $\tilde{\Psi}$ is considered to be dimensionless.
Using the quaternion product rule mentioned above and equating the real and imaginary components in the two sides of the resulting equation, yield
\begin{equation}
\vec{f}=-M\frac{\psi_0}{c}\,\frac{\vec{v}}{\tau}+\frac{M^2c^3}{\hbar}\,\vec{\psi}\,,\,\, P=\frac{Mc}{\tau}\,(\psi_0-\vec{v}\cdot\vec{\psi})\,,\,\, \vec{m}_t\cdot\vec{v}=m_\ell c\,,\,\, \tau=\frac{\hbar}{Mc^2}\,,
\end{equation}
upon employing eqs.(6) - (9).
The first term in the force is analogous to the electric force and the second one to the friction force that occurs in a conductor due to scattering between electrons. It is also like a viscous force that requires the presence of a fluid that permeates the whole space. The time $\tau$ is a characteristic time reflecting the jittery (trembling) motion of a quantum particle as observed by Schrodinger for the electron free-motion \textcolor[rgb]{0.00,0.00,1.00}{\cite{zitter}}. It seems a quantum particle jitters because it experiences a quantum (viscous) force  as it travels in space. The jitter frequency here is twice that due to the electron in the Dirac theory. The quantum resistive force  may explain the origin of inertia that any particle experiences as it starts to move.

\section{Concluding remarks}

We have extended in this work our earlier work on quantum mechanics  where the mass is treated as a quaternion. The quaternionic  momentum eigen-value equation leads to a Telegraph equation when a real mass is used, it however leads to the Klein-Gordon equation when a quaternionic mass is used instead. This scalar field has an effective mass that is a combination of its two masses.  The vector and scalar masses connect the particle matter-wave duality where the wave's momentum is equal to the particle's momentum.  This is expressed by the equation, $\vec{m}\,c=m_0\vec{v}$, where $\vec{v}$  is the particle's velocity. When the quantum particle interacts with an electromagnetic field, the particle's angular momentum is coupled to the magnetic field. A quantum frictional force acting on the moving particle is found to be proportional to its velocity.  The interaction of the quantum particle with the electromagnetic field can be obtained from that of a free particle with quaternionic mass provided one defines $c\,\vec{m}_t\equiv q\vec{A}$ and $m_\ell c^2\equiv q\varphi$. A new  linear momentum field is found to be associated with the spin of the electromagnetic field. A charged particle interacting with an electromagnetic field is found to be governed by a Schrodinger-like equation where the particle spin (orbital) angular momentum is coupled to the electromagnetic field.

\end{document}